\documentclass[cameraready]{Interspeech}
\usepackage{dsfont}
\usepackage{bm}
\usepackage{algorithm}
\usepackage{algpseudocode}
\usepackage{amsmath}
\usepackage{amssymb}
\usepackage{multirow}
\usepackage{wasysym}
\usepackage{booktabs}
\usepackage{siunitx}

\title{Decoding Order Matters in Autoregressive Speech Synthesis}

\author[]{Minghui}{Zhao}
\author[]{Anton}{Ragni}

\address{
    School of Computer Science, University of Sheffield, United Kingdom
}

\email{mzhao39@sheffield.ac.uk, a.ragni@sheffield.ac.uk}


\usepackage{comment}

\begin{document}

\maketitle

\begin{abstract}
    Autoregressive speech synthesis has traditionally followed a left-to-right order, yet generation order is a modelling choice. This paper investigates decoding order through a masked diffusion framework that enables arbitrary decoding orders at inference time. To isolate the effect of decoding order from inductive biases introduced by learned discrete encoders, we operate on scalar quantised Mel-spectrograms. Our results demonstrate that the left-to-right strategy is suboptimal for capturing complex acoustic dependencies. The reverse, right-to-left, order consistently outperforms the traditional approach, while the adaptive confidence-based strategy \texttt{top1} achieves the highest Mean Opinion Score (MOS) among all model outputs. Analysis reveals that the most effective orders maintain local clusters of consecutive frames to balance long-range temporal dependencies with local coherence. These results establish decoding order as a critical parameter for improving synthesis quality.
\end{abstract}

\section{Introduction}
Autoregressive (AR) generation has long been central to speech synthesis, from earlier systems such as Wavenet \cite{DBLP:journals/corr/OordDZSVGKSK16} and Tacotron 2 \cite{DBLP:journals/corr/abs-1712-05884} to recent approaches that represent speech as discrete tokens such as \cite{soundstream2021, encodec2022, wang2023neuralcodeclanguagemodels}. In these systems, speech is generated sequentially, always following a left-to-right order that mirrors the natural progression of speech.

From a modelling perspective, left-to-right generation is not necessarily optimal. Speech dependencies extend beyond a simple causal chain: pauses and emphasis often depend on global context, while coarticulation reflects interactions between both past and future phones. Even when the model is conditioned on the full text or phone sequence, the decoding order can influence how effectively these acoustic dependencies are captured. Exploring alternatives to left-to-right generation is therefore important not only for assessing the widely adopted and unchallenged convention, but also for improving synthesis quality.

Viewing decoding order as a modelling choice, consider a factorisation of the ground-truth distribution $p(\boldsymbol{y}_{1:T})$ for a length-$T$ sequence. Let $\mathcal{P} = \{B_1, \dots, B_I\}$ be a partition of the index set $\{1,\dots,T\}$ and let $S_I$ denote the permutations of $\{1, \dots, I\}$. For any $\boldsymbol{\sigma} \in S_I$, the chain rule yields 
\begin{equation}\label{eq:chain-rule-order}
    p(\boldsymbol{y}_{{1:T}}\mid \boldsymbol{\sigma}, \mathcal{P})
    = \prod_{i=1}^{I} p(\boldsymbol{y}_{B_{\boldsymbol{\sigma}(i)}} \mid \boldsymbol{y}_{B_{\boldsymbol{\sigma}(<i)}}),
\end{equation}
where $B_{\boldsymbol{\sigma}(i)}$ is the subset of indices revealed at step $i$. While \eqref{eq:chain-rule-order} is an exact identity for any $\boldsymbol{\sigma}$, the invariance no longer holds once the conditionals are approximated by a parameterised model $p_\theta$. Because different orders expose the model to different acoustic contexts, the resulting approximation of $p(\boldsymbol{y}_{1:T})$ becomes dependent on the chosen $\boldsymbol{\sigma}$. This work focuses on the one-by-one decoding regime where $|B_i|=1$ for all $i$ (thus $I=T$). Left-to-right decoding is the special case $B_{\boldsymbol{\sigma}(i)}=\{i\}$, but \eqref{eq:chain-rule-order} imposes no such constraint.
While separate AR models could be trained under different fixed orders to identify the optimal approximation, the $T!$ possible permutations make this computationally prohibitive and prevents comparison within a single unified framework.


\begin{figure}[!t]
  \centering
  \includegraphics[width=0.48\textwidth]{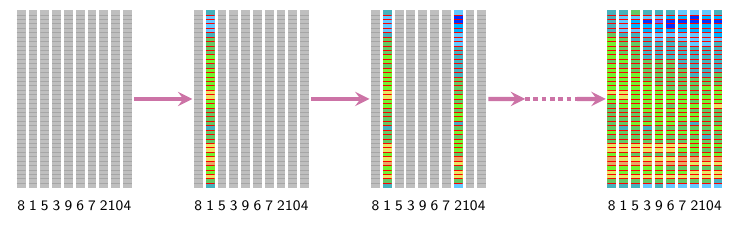}
  \caption{Intermediate Mel-spectrograms shown from left to right as generation progresses in a random order. The number beneath each frame indicates its order in generation and the rightmost frames are the final output.}
  \label{fig:ardm-inference}
\end{figure}

The masked diffusion model (MDM) \cite{sahoo2024simple,ardm,kim2025ordering} is a recently proposed framework that supports arbitrary decoding orders $\boldsymbol{\sigma} \in S_T$, thereby permitting AR factorisations under any permutation. During training, the model predicts randomly masked tokens from visible ones, so that each token is, in expectation, equally likely to be reconstructed. This \textit{order-agnostic} objective avoids favouring any particular information-revealing order. At inference (Fig.~\ref{fig:ardm-inference}), any generation order $\boldsymbol{\sigma}$ can be selected, unmasking positions sequentially conditioned on previously revealed tokens, allowing direct assessment of how order affects generation quality. To ensure that such analysis is not confounded by representation learning, we obtain tokens by directly discretising Mel-spectrograms via scalar quantisation.
Using this framework, we investigate how different decoding orders affect the quality of AR speech synthesis.
\section{Related work}
\subsection{Generative modelling framework}

Many recent works explore combining diffusion with autoregression for sequence generation. Block Diffusion \cite{blockdiff} and speech models such as DiffAR \cite{diffar} and DiTAR \cite{jia2025ditar} group consecutive frames into blocks and maintain a left-to-right AR generation across blocks, while predicting content within each block in parallel. We instead adopt the order-agnostic framework of \cite{ardm, kim2025ordering} and focus on the one-by-one regime.
The prior work most similar in method is MaskGCT \cite{wang2025maskgct}, which optimises a similar loss. Our objective, however, is explicitly designed to enable permutation-invariant generation. Beyond speech synthesis, alternative generation orders have been examined in reasoning \cite{kim2025ordering}, text generation \cite{BUSTOCASTINEIRA202589} and text-to-image \cite{fan2025fluid} settings, where the choice of $\boldsymbol{\sigma}$ affects performance, motivating exploration beyond left-to-right factorisations.

\subsection{Speech tokenisation}\label{sec:tokenisation}
Speech tokens are typically derived through learned quantisation \cite{soundstream2021, encodec2022} or clustering-based methods \cite{guo2025recentadvancesdiscretespeech}.
Because these discretisation mappings are trained under specific reconstruction or self-supervised objectives, they may introduce latent inductive biases to the resulting token inventory. 
As a result, analyses based on a particular learned tokeniser may not generalise to alternative or future tokenisation schemes.
To isolate the effect of decoding order from the influence of a learned encoder, we apply scalar quantisation directly to Mel-spectrograms. This signal-derived, deterministic, parameter-free approach ensures that the tokenisation remains a transparent reflection of the acoustic signal.
Scalar quantisation has previously been used for discrete generative modelling in the image domain, notably by FSQ-VQ \cite{mentzer2024finite}.
As demonstrated in Sec.~\ref{sec:quantisation}, these quantised Mel-spectrograms remain compatible with off-the-shelf HiFi-GAN \cite{hifigan} vocoders without the need for specialised retraining.

\section{Method}
Autoregressive sequence decoding can be characterised by two design choices, the order $\boldsymbol{\sigma}$ in which frames are generated and the update size $|B_i|$, i.e., how many elements are decoded at each step.
While many models impose a fixed schedule (e.g., left-to-right, one by one), the framework used here performs iterative masked prediction (Fig.~\ref{fig:ardm-inference}), producing a predictive distribution for all positions at each step. A subset of positions is then selected according to a decoding schedule and updated, and the process repeats until all tokens are generated. To study decoding order, we restrict the model to update one position at a time ($|B_i|=1$ for every step) without replacement, making it directly comparable to conventional AR models, as in \cite{ardm}.


The following sections describe inference and training. Let $\boldsymbol{\mu}$ denote a prior derived from text, with temporal length $T$. Conditioned on $\boldsymbol{\mu}$, the model generates an output sequence $\boldsymbol{y}$ of length $T$. In models such as Grad-TTS~\cite{gradtts} and Matcha-TTS~\cite{Mehta2023MatchaTTSAF}, all frames are updated simultaneously, whereas our model updates one frame at a time for $T$ steps.
Since MDMs operate on discrete data, Sec.~\ref{method:quantisation} details the quantisation method, while the motivation for this choice is discussed earlier in Sec.~\ref{sec:tokenisation}. Although decoding could be applied to individual time–frequency bins, we perform generation at the frame level. Sec.~\ref{sec:decoding-strategies} presents the decoding strategies. 

\subsection{Inference}
The inference procedure is outlined in Algorithm~\ref{alg:ardm-sampling}. 
While the decoding order $\boldsymbol{\sigma} \in S_T$ can follow any strategy, we illustrate a uniform random decoding, where $\boldsymbol{\sigma}$ is sampled uniformly from the permutation set $S_T$. Generation begins with all frames masked. For each generation step $t=1, \dots, T$ and position $i \in \{1, \dots, T\}$, we define the indicator masks
\begin{align}
\boldsymbol{m}_i &= \mathds{1}[\boldsymbol{\sigma}(i) < t], &
\boldsymbol{n}_i &= \mathds{1}[\boldsymbol{\sigma}(i) = t],
\end{align}
where $\boldsymbol{m}$ indicates previously generated frames and $\boldsymbol{n}$ the current position.
Conditioned on $\boldsymbol{\mu}$ and the context $\boldsymbol{m} \odot \boldsymbol{y}$, the model outputs parameters of a categorical distribution $\mathcal{C}(\boldsymbol{y} | \boldsymbol{f}_\theta(\boldsymbol{m} \odot \boldsymbol{y}, \boldsymbol{\mu}, \boldsymbol{m}))$, which approximates $p(\boldsymbol{y}_{\boldsymbol{\sigma}(t)} \mid \boldsymbol{y}_{\boldsymbol{\sigma}(<t)})$. An update $\boldsymbol{y}'$ is then drawn, with each frequency bin sampled independently, and only the frame indicated by $\boldsymbol{n}$ is updated. The process terminates when all $T$ frames are predicted, as illustrated in the spectrogram on the right side of Fig.~\ref{fig:ardm-inference}.

\begin{algorithm}[H]
\setlength{\textfloatsep}{5pt}
\caption{Order-agnostic Sampling}
\label{alg:ardm-sampling}
\begin{algorithmic}
\Require Conditioning input $\boldsymbol{\mu}$, Network $\boldsymbol{f}_\theta$
\Ensure Sample $\boldsymbol{y}$
\State Initialize $\boldsymbol{y} = \boldsymbol{0}$
\State Sample $\boldsymbol{\sigma} \sim \mathcal{U}(S_T)$
\For{$t$ in $\{1, \ldots, T\}$}
    \State $\boldsymbol{m} \leftarrow \bigl(\mathds{1}[\boldsymbol{\sigma}(i) < t]\bigr)_{i=1}^T, \boldsymbol{n} \leftarrow \bigl(\mathds{1}[\boldsymbol{\sigma}(i) = t]\bigr)_{i=1}^T$
    \State $\boldsymbol{y}' \sim \mathcal{C}(\boldsymbol{y} | \boldsymbol{f}_\theta(\boldsymbol{m} \odot \boldsymbol{y}, \boldsymbol{\mu}, \boldsymbol{m}))$
    \State $\boldsymbol{y} \gets (\boldsymbol{1} - \boldsymbol{n}) \odot \boldsymbol{y} + \boldsymbol{n} \odot \boldsymbol{y}'$
\EndFor
\end{algorithmic}
\end{algorithm}

\subsection{Order-agnostic training}\label{sec:training}
Algorithm~\ref{alg:ardm-optimizing} summarises the training procedure. Rather than maximising the chain-rule likelihood in Eq.~\eqref{eq:chain-rule-order}, which requires sequential prediction under a chosen permutation and is expensive when marginalising over permutations, we optimise the evidence lower bound (ELBO) on the log-likelihood \cite{ardm}:
\begin{align}
\label{eq:adrm-loss}
\log p(\boldsymbol{y}) &\ge T \cdot \mathbb{E}_{t, \boldsymbol{\sigma}} \left[ \frac{1}{T - t + 1} \sum_{k \in \boldsymbol{\sigma}(\ge t)} \log p(\boldsymbol{y}_k \mid \boldsymbol{y}_{\boldsymbol{\sigma}(<t)}) \right],
\end{align}
where $t \sim \mathcal{U}(1, \ldots, T)$ and $\boldsymbol{\sigma}$ is uniformly sampled from $S_T$.
The formulation in Eq.~\eqref{eq:adrm-loss} allows the conditional log-likelihood terms for all $k \in \boldsymbol{\sigma}(\ge t)$ to be computed in parallel for a sampled pair $(t, \boldsymbol{\sigma})$, yielding an efficient and order-agnostic objective. To approximate the expectation, we use a single-sample Monte Carlo estimate per step. 
\begin{algorithm}[H]
\setlength{\textfloatsep}{5pt}
\caption{Order-agnostic Training}
\label{alg:ardm-optimizing}
\begin{algorithmic}
\Require Data point $\boldsymbol{y}$, Conditioning input $\boldsymbol{\mu}$, Network $\boldsymbol{f}_\theta$
\Ensure $\mathcal{L}$
\State Sample $t \sim \mathcal{U}(1, \ldots, T)$
\State Sample $\boldsymbol{\sigma} \sim \mathcal{U}(S_T)$
\State Compute $\boldsymbol{m} \gets (\mathds{1}[\boldsymbol{\sigma}(i) < t])_{i=1}^T$
\State $\boldsymbol{l} \gets (\boldsymbol{1} - \boldsymbol{m}) \odot \log \mathcal{C}(\boldsymbol{y} |  \boldsymbol{f}_\theta(\boldsymbol{m} \odot \boldsymbol{y}, \boldsymbol{\mu}, \boldsymbol{m}))$
\State $\mathcal{L} \gets \frac{T}{T - t + 1} \text{sum}(\boldsymbol{l})$
\end{algorithmic}
\end{algorithm}


\subsection{Quantisation}\label{method:quantisation}
We apply a shared linear quantiser across frequency bins, yielding a parameter-free discretisation that avoids encoder-induced biases. Specifically, for a value $y \in [a,b]$, the quantised index $\hat{y}$ is given by:
\begin{equation}
\hat{y} = \operatorname{round}\left( \frac{y - a}{b-a} \cdot (Q - 1) \right),
\quad \hat{y} \in \{0, 1, \dots, Q-1\},
\end{equation}
where $Q$ denotes the number of quantisation bins.
Since the model predicts entire frames, each with $n_f$ bins taking one of $Q$ values, a frame corresponds to one of $Q^{n_f}$ possible tokens.


\subsection{Decoding strategies}\label{sec:decoding-strategies}


We examine fixed orders, left-to-right (\texttt{l2r}), and right-to-left (\texttt{r2l}) as well as adaptive top-$K$ and duration-guided strategies. These adaptive strategies prioritise high-confidence frames, inherently resulting in a non-linear generation path. To further explore this property, we evaluate stochasticity by interpolating between \texttt{l2r} and fully random (\texttt{uro}) orders. This allows us to observe how the model reacts to non-linearity, whether the specific decoding order is confidence driven or not.


\subsubsection{Top-$K$ probability}\label{sec:topk}
At step $t$, for each undecoded position $k\in\boldsymbol{\sigma}(\ge t)$ we compute
\begin{align}\label{eq:confidence-score}
    s_k(t)= \sum\limits_{i=1}^{n_f} \max_{j \in \{0,\dots,Q-1\}} \log p(\boldsymbol{y}_{k,i}=j\mid \boldsymbol{y}_{\boldsymbol{\sigma}(<t)}),
\end{align}
which sums the maximum log-probabilities across all $n_f$ bins. The next position is chosen as 
\begin{align}\label{eq:topk-selecting}
    k_t^\star \;=\; \arg\max_{k\in\boldsymbol{\sigma}(\ge t)} s_k(t), 
\end{align}
The decoding order is then updated as $\boldsymbol{\sigma}(t)\leftarrow k_t^\star$. In this work, we set $K=1$.
This strategy determines only decoding positions, not their values. We consider two variants, \texttt{top1}, which samples from the predicted distribution, and \texttt{top1*}, which selects the most probable values at each step.


\subsubsection{Duration-guided decoding}
While inspecting the adaptively determined orders (see Sec.~\ref{sec:decoding-order-impact}), we observed that the model often decodes contiguous positions, selecting adjacent frames consecutively. This suggests a preference for local continuity.
To examine whether chunk-wise decoding improves generation quality, we first segment each utterance. A simple approach is to use the duration predictor, which estimates how many frames each frame spans.
To select which segment to decode, we compute the average frame-level confidence $s_k(t)$ within each segment (Eq.~\eqref{eq:confidence-score}), and select the segment with the highest score; however, since the confidence is based on predicted probabilities, we note these probabilities may not be well calibrated \cite{estimating_confidence_scores}. The frames within the chosen segment are then updated in a random order, although different strategies could also be considered. This decoding scheme is denoted as \texttt{dur}.

\subsubsection{Stochastic path interpolation}\label{sec:control-randomness}
To evaluate the impact of non-linearity, we control the degree of randomness by perturbing the left-to-right order $\boldsymbol{\sigma}_{l2r}$ with swaps between two randomly chosen positions.
Theoretically, approximately $T \log T$ swaps produce a near-uniform permutation distribution \cite{kim2025ordering}. We scale this quantity by a factor $\beta \in (0, 1)$ to obtain a coarse measure of randomness, yielding orders close to\texttt{l2r} as $\beta \to 0$ and close to \texttt{uro} as $\beta \to 1$.


\section{Experiments}
\subsection{Experimental setup}
We use LJSpeech \cite{ljspeech17}, a corpus with $13,100$ clips (1-10s) of read speech by a single female native English speaker. We adopt the same split as Grad-TTS~\cite{gradtts}. As in \cite{gradtts, Mehta2023MatchaTTSAF}, the model consists of a text encoder, a duration predictor and a decoder. The encoder produces a latent text representation, upsampled using predicted durations to obtain $\boldsymbol{\mu}$, which serves as the decoding prior. We train the decoder as described in Sec.~\ref{sec:training}.
For each time-frequency bin, the decoder predicts parameters for a mixture of $5$ logistic components, sampled independently across Mel bins. At inference, a component is selected via the Gumbel-Max trick with temperature $t_1$ and the bin value is drawn from the chosen logistic distribution with temperature $t_2$.


We evaluate with Mel-Cepstral Distortion (MCD), $\log F_0$, UTMOSv2~\cite{baba2024utmosv2}, and Word Error Rate (WER),  all on 50 randomly sampled test-set utterances. WER is obtained using a HuBERT-based ASR model trained on read speech. \footnote{https://huggingface.co/facebook/hubert-large-ls960-ft} MCD and $\log F_0$ are calculated against vocoded references. For subjective evaluation, we report MOS via Amazon Mechanical Turk from 75 Master Workers (self-identified native English speakers), with 10 ratings per sample on average.\footnote{https://minghuizhao39.github.io/sample-page-order/}

\subsection{Effect of Mel-spectrogram quantisation}\label{sec:quantisation}
\begin{figure}[t]
    \centering
    \includegraphics[width=0.78\linewidth]
    {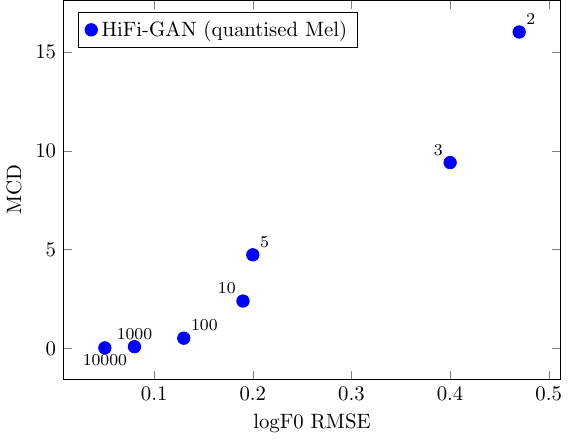}
    \vspace{-2mm}
    \caption{Effect of quantisation level on MCD and $\log F_0$ using a public checkpoint on scalar-quantised Mel (no retraining).}
    \label{fig:quantisation}
\end{figure}

Following Sec.~\ref{method:quantisation}, we conduct preliminary experiments using scalar quantisation of Mel-spectrograms. On the validation set, Mel features are discretised into $Q$ levels and vocoded with HiFi-GAN.
As expected, lower $Q$ increases MCD and $\log F_0$ (see Fig.~\ref{fig:quantisation}). At $Q=10$, both remain low ($2.40$ and $0.19$), indicating limited degradation relative to continuous Mel-spectrograms.
WER is already relatively low at $Q=5$ ($7.48$), decreases with larger $Q$, and stabilises at $5.22$ from $Q=100$ onward.
These results show that Mel-spectrograms are sufficiently redundant to support simple scalar quantisation while maintaining low MCD, $\log F_0$ and WER, and that standard vocoders such as HiFi-GAN can reconstruct high-quality speech from heavily quantised inputs. We therefore use $Q=100$ in subsequent experiments.


\subsection{Controlling randomness in decoding order}
Following Sec.~\ref{sec:control-randomness}, decoding order randomness is controlled by varying $\beta$: $0.001, 0.003, 0.01, 0.025, 0.1, 0.35, 0.5, 1.0$.
Smaller $\beta$ values yield orders close to \texttt{l2r}, whereas larger values approach fully mixed sampling (\texttt{uro}). To reduce variance due to random swaps, each $\beta$ setting is evaluated over $5$ independent runs and results are averaged per utterance.

\begin{figure}[t]
    \centering
    \includegraphics[width=0.85\linewidth,trim=0 0 0 0,clip]{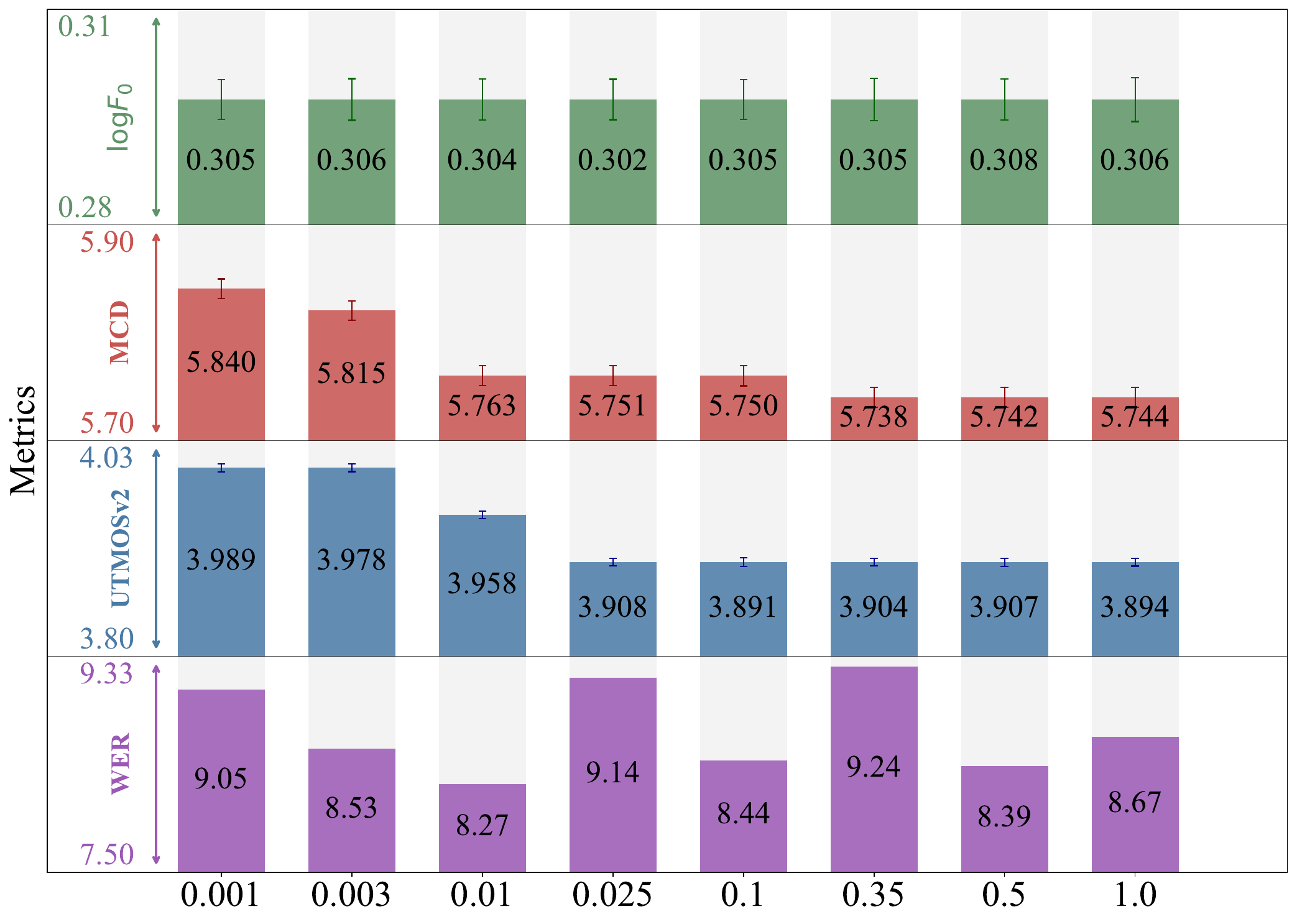}
    \vspace{-2mm}
    \caption{Evaluation across decoding orders with controlled randomness. Bars show metric values for different $\beta$.}
    \label{fig:randomness}
\end{figure}

As shown in Fig.~\ref{fig:randomness}, $\log F_0$ is largely insensitive to $\beta$. As $\beta$ increases, MCD decreases, indicating reduced spectral distortion under more mixed order. UTMOS, on the other hand, declines with increasing randomness, suggesting a preference for more sequential decoding. In contrast, WER exhibits non-monotonic behaviour, with a minimum at $\beta=0.01$.
Overall, the effect of decoding-order randomness on synthetic speech is metric-dependent. Although one might expect more left-to-right decoding to be advantageous, it does not consistently outperform more mixed orders.

\subsection{The impact of decoding order}\label{sec:decoding-order-impact}
We now evaluate specific decoding strategies directly.
For the baseline, we train Grad-TTS on full utterances instead of using a released checkpoint trained on segmented inputs ($172$ frames).
We evaluate two decoding strategies, fixed 100-step decoding (\texttt{grad-100}) and length-based decoding (\texttt{grad-L}). The former uses a sufficiently large number of steps to ensure high synthesis quality, while the latter matches the number of decoding steps used by the proposed model.

\begin{figure}[t]
    \centering
    \includegraphics[width=0.85\linewidth,trim=0 0 0 0,clip]{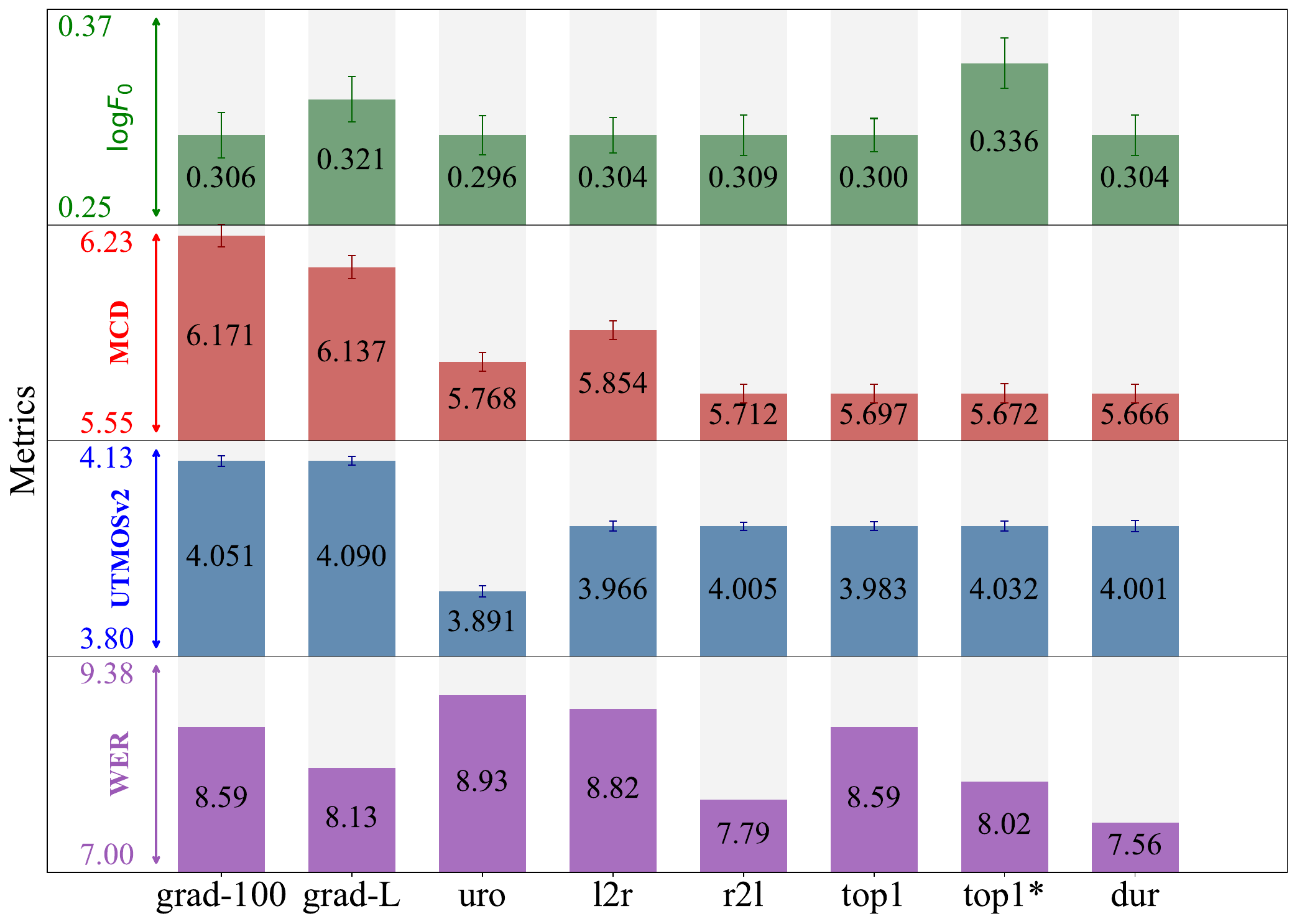}
    \vspace{-2mm}
    \caption{Evaluation on different decoding strategies}
    \label{fig:single-step}
\end{figure}

Automatic evaluation results are shown in Fig.~\ref{fig:single-step}. Overall, the adaptive strategy \texttt{dur} performs the best, closely followed by \texttt{r2l}, with low WER, MCD, $\log F_0$ and relatively high UTMOS (ground truth WER: $6.64$).
Among sequential strategies, \texttt{r2l} outperforms \texttt{l2r} in WER, UTMOS and MCD with comparable $\log F_0$, indicating that left-to-right decoding typical of AR models is not optimal. 
\texttt{top1*} performs similarly to \texttt{top1} in UTMOS and MCD. However, \texttt{top1} shows lower pitch distortion while \texttt{top1*} achieves better intelligibility. Deterministic argmax selection in \texttt{top1*} collapses each bin to its most probable value, reducing spectral variance across frequency bins. This loss of fine-grained variation may weaken harmonic structure and degrade pitch naturalness.
Although the Grad-TTS baselines generate continuous Mel-spectrograms, they yield marginal gains in UTMOS compared to other approaches but suffer from much higher MCD. Between the two models, \texttt{grad-L} shows worse $\log F_0$ despite improved WER.

MOS results show similar trend (Table~\ref{tab:mos}). Among systems, \texttt{top1} achieves the highest MOS after the vocoded reference, while \texttt{uro} performs worst.
\texttt{r2l} ranks second among the trained systems, outperforming some adaptive strategies and further suggesting that left-to-right decoding is suboptimal. 
The gap between \texttt{top1} and \texttt{top1*} likely reflects higher pitch distortion in \texttt{top1*}, which reduces perceived naturalness despite better intelligibility. \texttt{dur} shows only mediocre performance; the benefit of confidence-based, segment decoding appears to be offset by the random order within each segment. \texttt{l2r} and \texttt{top1*} are slightly worse than the Grad-TTS baselines.

Analysis shows that confidence-based decoding frequently generates consecutive frames in a right-to-left order. We quantify this using the right-to-left transition ratio $\rho_{r2l}$, defined as the proportion of steps satisfying $\boldsymbol{\sigma}(t-1)-\boldsymbol{\sigma}(t)=1$ (normalised by sequence length minus one); $\rho_{l2r}$ is defined analogously. For \texttt{top1}, $\rho_{r2l} \approx 0.9$ and $\rho_{l2r} \approx 0.05$, indicating decoding dominated by local right-to-left expansions. The strong subjective performance of \texttt{top1}, with the competitive results of \texttt{r2l}, suggests that locally right-to-left progression is associated with improved perceptual quality. The weaker performance of \texttt{dur}, which breaks this directional continuity through random ordering within segments, is consistent with this observation.
While this does not imply causality, it indicates a directional pattern correlated with better performance.


\begin{table}
  \caption{MOS for various decoding strategies. (95\% CI, Student's $t$ distribution, all ranging from $\pm 0.08$ to $\pm 0.09$).
 }
  \label{tab:mos}
  \footnotesize
  \centering
  \begin{tabular}{ c c c }
    \toprule
    \textbf{Order Type} & \textbf{Decoding Strategy} & \textbf{MOS} \\
    \midrule
    \multirow[c]{3}{*}{Baseline}
    & $\texttt{vocoded reference}$        & $3.99$~~~              \\
    & $\texttt{grad-100}$                 & $3.80$~~~              \\
    & $\texttt{grad-L}$                   & $3.81$~~~              \\
    \midrule
    \multirow[c]{3}{*}{Fixed}
    & $\texttt{uro}$                      & $3.50$~~~              \\
    & $\texttt{l2r}$                      & $3.78$~~~              \\
    & $\texttt{r2l}$                      & $\textbf{3.87}$~~~              \\
    \midrule
    \multirow[c]{3}{*}{Adaptive}
    & $\texttt{top1}$                     & $\textbf{3.91}$~~~              \\
    & $\texttt{top1*}$                    & $3.79$~~~              \\
    & $\texttt{dur}$                      & $3.81$~~~              \\
    \bottomrule
  \end{tabular}

\end{table}



\section{Conclusion and discussion}
This study shows that the conventional left-to-right decoding order in AR speech synthesis is suboptimal for capturing complex acoustic dependencies. Using an order-agnostic MDM, we find that strategies such as right-to-left and adaptive confidence-based decoding improve both objective and subjective metrics, with \texttt{top1} achieving the highest perceived naturalness.
Our results indicate that decoding order is a critical yet overlooked modelling factor. Performance improves when decoding balances global context with local coherence, particularly through contiguous frame generation with a locally dominant right-to-left progression. 
Scalar-quantised Mel-spectrograms ensure that these improvements stem from decoding order rather than encoder-specific token representations.
Experiments also show substantial redundancy in Mel-spectrograms, as high-quality speech can be reconstructed from heavily quantised inputs.
Future work will examine whether these findings generalise to multi-speaker systems or more diverse acoustic conditions.

\section{Generative AI Use Disclosure}
Generative AI tools were used to assist with the creation of Fig.~\ref{fig:ardm-inference}, the development of the audio demo webpage, and polishing the manuscript.

\ifcameraready{}
\section{Acknowledgement}
This work was supported by the UKRI AI Centre for Doctoral Training in Speech and Language Technologies (SLT) and their Applications funded by UK Research and Innovation [grant number EP/S023062/1]. For the purpose of open access, the author has applied a Creative Commons Attribution (CC BY) licence to any Author Accepted Manuscript version arising.
\else
\fi

\bibliographystyle{IEEEtran}
\bibliography{mybib}

\end{document}